\documentclass[sigconf]{acmart}
\pdfoutput=1

\makeatletter                   
\def\mdseries@tt{m}             
\makeatother                    
\usepackage[plain]{fancyref}
\usepackage[draft=true]{minted} 

\usepackage{hyperref}

\usepackage[utf8]{inputenc}
\usepackage{multirow}
\usepackage{pbox}
\usepackage{diagbox, eqparbox, hhline}
\usepackage{pgf}
\usepackage{xspace}
\usepackage{xcolor}
\usepackage{booktabs}
\usepackage{amssymb}
\usepackage{amsmath}
\usepackage{pgfplots}\pgfplotsset{compat=1.9}
\usepackage{pgfplotstable}
\usepackage{booktabs,colortbl}

\usepackage{graphicx}

\DeclareMathOperator*{\argmax}{arg\,max}

\makeatletter
\renewcommand\@formatdoi[1]{\ignorespaces}
\makeatother
\setcopyright{rightsretained}
\acmDOI{ }
\acmISBN{}
\setcopyright{rightsretained}
\acmConference[RecSysKTL'18]{Workshop on Intelligent Recommender Systems by Knowledge Transfer and Learning}{October 2018} {Vancouver, Canada}
\acmYear{2018}
\copyrightyear{2018}
\acmArticle{}
\acmPrice{}

\begin{document}
\newcommand{\knn}{$k$-NN\xspace} 				
\newcommand{\neighOR}{neighbor\xspace} 			
\newcommand{\neighsOR}{neighbors\xspace} 		
\newcommand{\neighhoodOR}{neighborhood\xspace} 	
\newcommand{\NeighhoodOR}{Neighborhood\xspace} 	
\newcommand{\NeighOR}{Neighbor\xspace} 			
\newcommand{\NeighsOR}{Neighbors\xspace} 		
\newcommand{\checkins}{check-ins\xspace}
\newcommand{\checkin}{check-in\xspace}
\newcommand{\Checkins}{Check-ins\xspace}
\newcommand{\Checkin}{Check-in\xspace}

\newcommand{\Istanbul}{Istanbul\xspace}
\newcommand{\Jakarta}{Jakarta\xspace}
\newcommand{\KualaLumpur}{Kuala Lumpur\xspace}
\newcommand{\Mexico}{Mexico City\xspace}
\newcommand{\Moscow}{Moscow\xspace}
\newcommand{\Santiago}{Santiago\xspace}
\newcommand{\SaoPaulo}{S\~{a}o Paulo\xspace}
\newcommand{\Tokyo}{Tokyo\xspace}

\newcommand{\IstanbulAbbr}{IST\xspace}
\newcommand{\JakartaAbbr}{JAK\xspace}
\newcommand{\KualaLumpurAbbr}{KUA\xspace}
\newcommand{\MexicoAbbr}{MEX\xspace}
\newcommand{\MoscowAbbr}{MOS\xspace}
\newcommand{\SantiagoAbbr}{SAN\xspace}
\newcommand{\SaoPauloAbbr}{SAO\xspace}
\newcommand{\TokyoAbbr}{TOK\xspace}

\newcommand{\geomf}{GeoMF\xspace}
\newcommand{\igslr}{iGSLR\xspace}
\newcommand{\gscorr}{gSCorr\xspace}
\newcommand{\lrt}{LRT\xspace}
\newcommand{\irenmf}{IRenMF\xspace}
\newcommand{\rankgeofm}{Rank-GeoFM\xspace}
\newcommand{\lfbca}{LFBCA\xspace}

\newcommand{\AvgDist}{AvgDis\xspace}
\newcommand{\PopGeoNN}{PGN\xspace}

\newcommand{\FoldNov}{Nov\xspace}
\newcommand{\FoldMay}{May\xspace}

\newcommand{\Domain}{\mathcal{D}}
\newcommand{\Users}{\mathcal{U}}
\newcommand{\Items}{\mathcal{L}}

\newcommand{\sep}{1.7cm}
\newcommand{\sepParams}{2.5cm}
\newcommand{\fontTables}{\normalsize}

\newcommand{\blacktriangleSmall}{\text{\tiny\ensuremath\blacktriangle}}
\newcommand{\blacktriangledownSmall}{\text{\tiny\ensuremath\blacktriangledown}}
\newcommand{\daggerSmall}{\text{\tiny\ensuremath\dagger}}

\newcommand{\findmax}[4]{
    \pgfplotstablevertcat{\datatable}{#1}%
    \pgfplotstablecreatecol[
    create col/expr={%
    \pgfplotstablerow
    }]{rownumber}\datatable%
    \pgfplotstablesort[sort key={#2},sort cmp={float >}]{\sorted}{\datatable}%
    \pgfplotstablegetelem{0}{rownumber}\of{\sorted}%
    \pgfmathtruncatemacro#3{\pgfplotsretval+#4}%
    \pgfplotstableclear{\datatable}%
}
\newcommand{\findmin}[4]{
    \pgfplotstablevertcat{\datatable}{#1}
    \pgfplotstablecreatecol[
      create col/expr={%
    \pgfplotstablerow
    }]{rownumber}\datatable
    \pgfplotstablesort[sort key={#2},sort cmp={float <}]{\sorted}{\datatable}%
    \pgfplotstablegetelem{0}{rownumber}\of{\sorted}%
    \pgfmathtruncatemacro#3{\pgfplotsretval+#4}
    \pgfplotstableclear{\datatable}
}
\pgfplotstableset{
    highlight col maxP/.code n args={4}{
        \pgfplotstabletranspose{\datamaxP}{#1}
        \pgfplotstabletranspose[columns=#4, colnames from=colnames]{\datamaxPTrans}{\datamaxP}
        \findmax{\datamaxPTrans}{#2}{\maxval}{#3}
        \edef\setstyles{\noexpand\pgfplotstableset{
                every row \maxval\noexpand\space column #2/.style={
                    postproc cell content/.append style={
                        /pgfplots/table/@cell content/.add={$\noexpand\bf}{$}
                    },
                }
            }
        }\setstyles
    },
    highlight col maxPG/.code n args={4}{
        \pgfplotstabletranspose{\datamaxP}{#1}
        \pgfplotstabletranspose[columns=#4, colnames from=colnames]{\datamaxPTrans}{\datamaxP}
        \findmax{\datamaxPTrans}{#2}{\maxval}{#3}
        \edef\setstyles{\noexpand\pgfplotstableset{
                every row \maxval\noexpand\space column #2/.style={
                    postproc cell content/.append style={
                        /pgfplots/table/@cell content/.add={$\noexpand\daggerSmall\noexpand\bf}{$}
                    },
                }
            }
        }\setstyles
    },
    highlight col max/.code n args={3}{
        \findmax{#1}{#2}{\maxval}{#3}
        \edef\setstyles{\noexpand\pgfplotstableset{
                every row \maxval\noexpand\space column #2/.style={
                    postproc cell content/.append style={
												/pgfplots/table/@cell content/.add={$\noexpand\dagger$}{}
                    },
                }
            }
        }\setstyles
    },
    highlight col min/.code n args={3}{
        \findmin{#1}{#2}{\minval}{#3}
        \edef\setstyles{\noexpand\pgfplotstableset{
                every row \minval\noexpand\space column #2/.style={
                    postproc cell content/.append style={
                        /pgfplots/table/@cell content/.add={\noexpand\color{red}$\noexpand\bf}{$}
                    },
                }
            }
        }\setstyles
    },
    highlight row max/.code n args={4}{
        \pgfmathtruncatemacro\rowindex{#2-1}
        \pgfplotstabletranspose[columns=#3]{\transposed}{#1}
        \findmax{\transposed}{\rowindex}{\maxval}{#4}
        \edef\setstyles{\noexpand\pgfplotstableset{
                every row \rowindex\space column \maxval\noexpand/.style={
                    postproc cell content/.append style={
                        /pgfplots/table/@cell content/.add={$\noexpand\bf}{$}
                    },
                }
            }
        }\setstyles
    },
    highlight row maxT/.code n args={4}{
        \pgfmathtruncatemacro\rowindex{#2-1}
        \pgfplotstabletranspose[columns=#3]{\transposed}{#1}
        \findmax{\transposed}{\rowindex}{\maxval}{#4}
        \edef\setstyles{\noexpand\pgfplotstableset{
                every row \rowindex\space column \maxval\noexpand/.style={
                    postproc cell content/.append style={
                        /pgfplots/table/@cell content/.add={$\noexpand\blacktriangleSmall$}{}
												%
                    },
                }
            }
        }\setstyles
    },
		rowreducedecs/.style={
			postproc cell content/.append code={
				\count0=\pgfplotstablerow
				\advance\count0 by1
				\ifnum\count0=#1
					\pgfkeysalso{@cell content=\textit{\pgfmathprintnumber[fixed,precision=1,fixed zerofill]{##1}}}%
				\fi
			},
		},
    highlight row min/.code n args={4}{
        \pgfmathtruncatemacro\rowindex{#2-1}
        \pgfplotstabletranspose[columns=#3]{\transposed}{#1}
        \findmin{\transposed}{\rowindex}{\minval}{#4}
        \edef\setstyles{\noexpand\pgfplotstableset{
                every row \rowindex\space column \minval\noexpand/.style={
                    postproc cell content/.append style={
                        /pgfplots/table/@cell content/.add={$\noexpand\it}{$}
                    },
                }
            }
        }\setstyles
    },
    highlight row minT/.code n args={4}{
        \pgfmathtruncatemacro\rowindex{#2-1}
        \pgfplotstabletranspose[columns=#3]{\transposed}{#1}
        \findmin{\transposed}{\rowindex}{\minval}{#4}
        \edef\setstyles{\noexpand\pgfplotstableset{
                every row \rowindex\space column \minval\noexpand/.style={
                    postproc cell content/.append style={
                        /pgfplots/table/@cell content/.add={$\noexpand\blacktriangledownSmall$}{}
                    },
                }
            }
        }\setstyles
    },
}


\title{
A novel approach for venue recommendation using cross-domain techniques
}

\author{Pablo S\'{a}nchez}
\orcid{0000-0003-1792-1706}
\affiliation{%
	\institution{Universidad Aut\'onoma de Madrid}
	\city{Madrid}
	\postcode{28049}
	\country{Spain}
}
\email{pablo.sanchezp@uam.es}
\author{Alejandro Bellog\'{i}n}
\orcid{0000-0001-6368-2510}
\affiliation{%
	\institution{Universidad Aut\'onoma de Madrid}
	\city{Madrid}
	\postcode{28049}
	\country{Spain}
}
\email{alejandro.bellogin@uam.es}

\begin{abstract}
Finding the next venue to be visited by a user in a specific city is an interesting, but challenging, problem.
Different techniques have been proposed, combining collaborative, content, social, and geographical signals; however it is not trivial to decide which technique works best, since this may depend on the data density or the amount of activity logged for each user or item.
At the same time, cross-domain strategies have been exploited in the recommender systems literature when dealing with (very) sparse situations, such as those inherently arising when recommendations are produced based on information from a single city.

In this paper, we address the problem of venue recommendation from a novel perspective: applying cross-domain recommendation techniques considering each city as a different domain.
We perform an experimental comparison of several recommendation techniques in a temporal split under two conditions: single-domain (only information from the target city is considered) and cross-domain (information from many other cities is incorporated into the recommendation algorithm). 
For the latter, we have explored two strategies to transfer knowledge from one domain to another: testing the target city and training a model with information of the $k$ cities with more ratings or only using the $k$ closest cities. 

Our results show that, in general, applying cross-domain by proximity increases the performance of the majority of the recommenders in terms of relevance.
This is the first work, to the best of our knowledge, where so many domains (eight) are combined in the tourism context where a temporal split is used, and thus we expect these results could provide readers with an overall picture of what can be achieved in a real-world environment.
\end{abstract}

\maketitle

\section{Introduction}
\label{sec:intro}
	The great development of location-based social networks (LBSNs) in recent years has encouraged the research into the problem of Point-of-Interest (POI) or venue recommendation, i.e., suggesting new places to visit by analyzing the users' tastes, needs, and movement patterns. Foursquare, Gowalla, or GeoLife are examples of these kind of social networks, where users record \checkins they make to certain venues (restaurants, cinemas, hotels, etc.) and share their experiences with other users in the system \citep{DBLP:conf/gis/ZhangC13,DBLP:conf/sigir/YeYLL11}. This information, if processed and exploited correctly, can be then used to suggest new venues to visit when using a recommendation engine.
	
Since research on Recommender Systems (RS) has increased in many directions in the last years, it is important to consider some specific details of POI recommendation that differ from the more traditional recommendation problem \citep{DBLP:conf/sigir/LiCLPK15,DBLP:conf/gis/WangTM13,DBLP:journals/pvldb/LiuPCY17}. 
These include, but are not limited, to:

\begin{itemize}
	\item \textbf{Sparsity:} in the RS domain, the user-item matrix is normally very sparse. However, in venue recommendation, this effect is even more severe. For example, for the Movielens20M and Netflix datasets their densities are $0.539\%$ and $1.177\%$, respectively. For the dataset we shall be using in our experiments (from Foursquare), the density is $0.0034\%$.
	\item \textbf{Implicit information:} \checkin data is one type of implicit feedback, where only positive values indicating that a user has visited a venue are recorded. 
	Nonetheless, since users may \checkin in the same place several times, researchers often build frequency matrices to model these repetitions. This differs from traditional recommendation, where it is normally assumed that users rate each item once \citep{DBLP:reference/sp/NingDK15}.	
	\item \textbf{External influences:} venue recommendation is highly affected by social (user friends), temporal, and geographical influence. The latter is possibly the most important effect to take into account in POI recommendation to improve the recommendation performance as it is usually assumed that users prefer to visit venues that are close to each other (as the first law of geography states ``\textit{Everything is related to everything else, but near things are more related than distant things}'' \cite{MHJ2004}). That is the reason why researchers have proposed many algorithms including some component of geographical influence \citep{DBLP:conf/cikm/LiuWSM14,DBLP:conf/sigir/YeYLL11,DBLP:conf/kdd/LianZXSCR14}. 
	
	
\end{itemize}

\iffalse
A large number of techniques specifically tailored for venue recommendation have been proposed that consider these characteristic aspects of the problem.
Some of them are based on matrix factorization techniques like LRT \citep{DBLP:conf/recsys/GaoTHL13} and \irenmf \citep{DBLP:conf/cikm/LiuWSM14}, and some others use classic collaborative filtering similarities like USG \citep{DBLP:conf/sigir/YeYLL11} and iGSLR \citep{DBLP:conf/gis/ZhangC13} (all of them incorporating the geographical influence in several ways). 
	
On the other hand, different techniques can be explored to obtain or access more information in order to help recommenders with their suggestions. 
\else
At the same time, different techniques can be explored to obtain or access more information in order to help recommenders with their suggestions. 
\fi
Transfer (or cross-domain) learning is one of these valuable techniques that allow to make use of external or additional information, and integrate it with current knowledge. 
In the context of RS, cross-domain recommendation is a recent and active research topic, where tourism has been acknowledged as a potential target domain~\cite{DBLP:journals/tbd/Zheng15}. 
However, not many experimental comparisons have been performed using cross-domain techniques in the tourism field with a realistic evaluation, especially exploiting geographical or social influence.
	
Thus, based on the fact that there has not been an exhaustive analysis of the effect of different cross-domain strategies in venue recommendation, in this paper we analyze the effect of producing recommendations by using only the venues of each separate city (single-domain) and contrast it against the combination of information extracted from other cities (cross-domain). 
Therefore, in this paper we address the following research questions:

\begin{description}
\item[RQ1:] \textbf{Are state-of-the-art recommendation algorithms able to exploit cross-domain information in the context of venue recommendation?} We empirically compare a set of recommenders under two conditions: single-domain (only information from the target city is considered) and cross-domain (information from many other cities is incorporated into the recommendation algorithm). We further explore different ways to select other cities to learn from in the cross-domain condition -- more specifically, according to their popularity and the distance to the target city.
\item[RQ2:] \textbf{Which is the best cross-domain strategy in terms of relevance?} We focus on nDCG as ranking metric and see which cross-domain strategy has a larger impact in performance with respect to the single-domain approach.
\end{description}

Our work provides a thorough comparison of two cross-domain strategies for venue recommendation under a realistic evaluation by means of a time-aware methodology. We compare the results obtained by the cross-domain strategies against the single-domain scenario in terms of ranking accuracy using a real dataset with more than $30$M interactions.
Our results indicate that cross-domain strategies can improve the performance of some recommendation approaches, whereas for other recommendation algorithms -- such as those based on geographical information -- single-domain scenarios seem to be more appropriate.

\section{Background}
\label{sec:background}
In the RS community, a large number of algorithms have been proposed in order to learn and model the user interests. 
The most well-known approaches are collaborative filtering (CF) and content-based (CB) algorithms, which are usually combined, creating hybrid RS. 
However, given their pervasiveness in the literature, we will skip their definitions here and refer the reader towards \cite{DBLP:reference/sp/GemmisLMNS15} to learn about CB systems and \cite{DBLP:reference/sp/NingDK15,DBLP:reference/sp/KorenB15} to know more about CF techniques, such as nearest neighbors and matrix factorization approaches. 
Hence, in this section we first focus on the specific aspects of venue recommendation (Section~\ref{sec:poi-rec}) and later we present the main concepts about cross-domain recommendation (Section~\ref{sec:cross-domain}).

\subsection{Venue recommendation}
\label{sec:poi-rec}
As already mentioned, venue suggestion is related to the traditional recommendation problem, even though it exhibits some distinctive features. 
Because of this, we introduce now some specific notation for this task, following the one presented in \cite{DBLP:journals/pvldb/LiuPCY17}. Let us denote the set of users with $\Users$ and the set of venues (or POIs) with $\Items$. We use $\Items^u$ to indicate the venues visited by user $u$ in the training set. The main objective is to recommend venues that the user has not seen beforehand, even though repetitions can be exploited when building the recommendation model. 

This type of recommendation is highly influenced by spatial, social, and temporal factors \cite{DBLP:journals/pvldb/LiuPCY17}, so most of the algorithms try to model them in order to perform recommendations.
Matrix factorization (MF) techniques are one of the most extended approaches in the RS area because they tend to outperform other techniques such as neighbor-based algorithms \cite{DBLP:journals/computer/KorenBV09}, they can also be easily extended with baseline predictors and temporal information \cite{DBLP:reference/sp/KorenB15}. 
In venue recommendation they are also very popular and they are the core of many approaches, such as LRT \citep{DBLP:conf/recsys/GaoTHL13}, \irenmf \citep{DBLP:conf/cikm/LiuWSM14}, and others \cite{DBLP:conf/kdd/LianZXSCR14,DBLP:conf/sigir/LiCLPK15}. 
As in classical recommendation, the basic operation in these algorithms is the minimization of the following function:
\begin{equation}
	\min_{\Users,\Items} || C - \hat{C} ||^{2}_F + \lambda_1||\Users||^{2}_F + \lambda_2||\Items||^2_F
\end{equation}
	where $\hat{C}$ is computed by the product of the user and POI matrices $\hat{C}=\Users\Items^T$, $F$ is the Frobenius norm, and $\lambda_1$ and $\lambda_2$ are the regularization parameters. 
	This is the basic algorithm, but it has been extended and adapted to this task. 
	For example, the \irenmf approach \cite{DBLP:conf/cikm/LiuWSM14} adds the geographical influence by using the $k$ nearest \neighsOR of the target venue in the score function, where a clustering based on the POIs locations is also applied. 
	The LRT algorithm \cite{DBLP:conf/recsys/GaoTHL13} includes temporal information in the MF algorithm by introducing temporal steps to factorize the \checkin matrix by each time slot; it then aggregates the result and adds a regularization term based on the similarities between the \checkin matrix at different points in time. 
	Other techniques using MF approximations at the core of their algorithms include the \geomf \cite{DBLP:conf/kdd/LianZXSCR14} and the \rankgeofm \cite{DBLP:conf/sigir/LiCLPK15} approaches.

However, other methods besides those based on MF have also been applied in venue recommendation. 
For example, the USG model from \cite{DBLP:conf/sigir/YeYLL11} combines three different components: a probabilistic model based on the history of visited POIs by the user to consider geographical information, CF similarities based on other users in the system (classic CF) , and CF similarities based on the friends of the target user (social influence). 
Another interesting approach is the LORE algorithm from \cite{DBLP:conf/gis/ZhangCL14}, where Markov Chains are used to model the sequential patterns between POIs, together with social and geographical influence. 
As can be seen, there is a great number of venue recommendation algorithms but the basis for most of them are proposals already explored in the traditional recommendation problem, complementing them with specific features of LBSNs (geographical, social, and temporal influence).
In this paper, we propose to exploit cross-domain strategies in this context; in the next section we define the cross-domain problem and explain its main concepts.

\subsection{Cross-domain recommendation}
\label{sec:cross-domain}
In the context of cross-domain recommendation, there exist multiple definitions of ``domain'' \cite{DBLP:reference/sp/CantadorFBC15}: one can consider two items belong to different domains if they have different values for a specific attribute but others may argue different domains implies two separate systems or two types of items. 
In any case, the basic idea behind this type of recommendation is that, to improve recommendations over a target domain $\Domain_T$, some kind of knowledge from a source domain $\Domain_S$ needs to be exploited. 
In order to illustrate these concepts, consider for example a source domain of books and a target domain of movies. We can make recommendations of users who do not have a mature rating history in the target domain by using some information stored in the source domain, i.e., if the target user liked books of drama and terror, she would probably also like movies of the same style even if the items are from different domains.
	
Depending on the information analyzed and the destination users to make recommendations, in \cite{DBLP:reference/sp/CantadorFBC15} the following three main recommendation goals are described:
\begin{itemize}
	\item Linked-domain recommendations: recommend items in the target domain by analyzing both the target and source domains.
	\item Cross-domain recommendations: recommend items in the target domain to users in the source domain by using only the information of the source domain.
	\item Multi-domain recommendations: recommend items belonging to the target or the source domain to all kind of users. 
\end{itemize}

Furthermore, we must consider that $\Domain_S$ and $\Domain_T$ may share some information about the users, the items, or both, which allows us to categorize the different scenarios according to the data overlap as: no overlap, item overlap, user overlap, and full overlap (both user and item overlap). 
Additionally, we can also classify the cross-domain techniques according to how they exploit the knowledge. 
Based on \cite{DBLP:reference/sp/CantadorFBC15}, there are two different categories: ``aggregating knowledge'', when the knowledge of the domains (user preferences, similarities, or single-domain recommendations) is aggregated, and ``linking and transferring knowledge'', where there is a knowledge transfer between the different domains in order to produce recommendations, for instance by using common knowledge (semantic networks, items attributes, etc), sharing latent features, or transferring the rating patterns.

In Section~\ref{sec:soa} we survey some proposals where cross-domain techniques were used for venue recommendation and related tasks.
Nonetheless, we want to emphasize we have not found many examples of cross-domain experiments combining more than two or three domains -- usually, movies, music, and books --, except for the works presented in \cite{DBLP:conf/cikm/RafailidisC17} and \cite{DBLP:conf/recsys/SahebiB15}. In the first one, the authors exploited the information of $10$ domains (categories of different products from Epinions, hence, not related with tourism) in order to analyze the performance of the recommendations using ranking metrics (nDCG and Recall), however, these datasets are much smaller in terms of ratings than the ones we use in this paper (the largest one contains around $200$K ratings) and they are all more dense; while in the second one, the authors used $21$ domains (defined as the categories from different Yelp businesses) and measured the effects of cross-domain recommendation in terms of RMSE by comparing the performance on several domain pairs. 
Therefore, our work is -- to best of our knowledge -- the first study where each city defines a separate domain to exploit cross-domain strategies, using up to $8$ domains to be combined into the source domain, and in the context of a tourism dataset with a realistic, temporal ranking-based evaluation.

\section{Adapting cross-domain recommendation to venue suggestion}
\label{sec:crosspoi}
Taking into account the characteristics of cross-domain recommendation, we propose to apply and adapt these concepts to venue recommendation. 
In LBSNs, it is natural to group the \checkins provided by the users per city, this, together with the issue that there is no venue overlap between different cities, has caused that many researchers have isolated the \checkins of different cities and considered them as independent datasets~\cite{DBLP:conf/sigir/LiCLPK15,DBLP:journals/tois/LiJHL17, DBLP:conf/cikm/LiuWSM14}.

In this paper, we consider each city as a different domain and propose to adapt strategies from cross-domain recommendation to combine these domains.
The main advantage of applying some kind of cross-domain in venue recommendation is that we can expand the knowledge of the recommenders with a larger number of users and items, in order to establish more relationships between them. 
However, it is not obvious how the knowledge to be included should be selected, since noise might be added to the model or, simply, that information may not to be useful at all.

The main goal we aim to achieve, if we focus on CF algorithms, would be to find highly active users in the aggregated domains so that they could alleviate the inherent sparsity problem.
Therefore, we consider the following possibilities:
\begin{itemize}
	\item Nearest cross-domain (N-CD): we use the $n$ closest cities to the target city as the source domain. We aim to capture cultural patterns~\cite{DBLP:journals/tist/YangZQ16}, while, at the same time, we keep control of the number of cities we consider. 
	\item Most-popular cross-domain (P-CD): we use the $n$ cities with more ratings as the source domain. This strategy allows us to test whether considering those cities that the system has more information about can be useful for the model. We hypothesize that having more information should be helpful for the recommendation algorithms, however, this strategy might also be more sensitive to noise and may not improve the user overlap.
\end{itemize}

\begin{table*}[tb]
\fontTables
\centering
\caption{Average percentage of common users with respect to the other cities included in each cross-domain strategy.}
\label{t:usersAverage}
\vspace{-10pt}
\iffalse
\begin{tabular}{crr}
\toprule
\textbf{City} & \textbf{N-CD} & \textbf{P-CD} \\ 
\midrule
\IstanbulAbbr & 9.36\%                         & 0.33\%                        \\ 
\JakartaAbbr & 13.58\%                        & 1.10\%                        \\ 
\KualaLumpurAbbr & 28.77\%                       & 0.90\%                        \\ 
\MexicoAbbr & 11.61\%                       & 0.61\%                        \\ 
\midrule
\MoscowAbbr & 2.82\%                        & 0.36\%                        \\ 
\SantiagoAbbr & 13.07\%                       & 0.74\%                        \\ 
\SaoPauloAbbr & 7.43\%                        & 0.62\%                        \\ 
\TokyoAbbr & 33.26\%                       & 0.49\%                       \\
\bottomrule
\end{tabular}
\else
\begin{tabular}{crrrrrrrr}
\toprule
Cross-domain & \multicolumn {8}{c}{Cities} \\ 
Strategy & \Istanbul &	\Jakarta & \KualaLumpur &	\Mexico &	\Moscow &	\Santiago &	\SaoPaulo &	\Tokyo \\
\midrule
N-CD & 9.36\%		& 13.58\%		& 28.77\%		& 11.61\%			& 2.82\%		& 13.07\%		& 7.43\%		& 33.26\% \\
P-CD & 0.33\%		& 1.10\%		& 0.90\%		& 0.61\%			& 0.36\%		& 0.74\%		& 0.62\%		& 0.49\% \\
\bottomrule
\end{tabular}
\fi
\vspace{-10pt}
\end{table*}

The rationale behind our proposal is exposed in Table~\ref{t:usersAverage}.
Here we show the average of the percentage of common users between a given target city against the other cities that belong to the source domain, i.e., either the closest cities for N-CD or most popular cities for P-CD.
Using the same dataset that will be used (and further explained) later in the experiments (prior partitioning and other processing), we compute the percentage of common users as follows (an analysis consistent with the one presented in~\cite{DBLP:journals/tist/YangZQ16}):

\begin{equation}
	\mbox{Common Users}(C_1, C_2) = \frac{| \Users(C_1) \cap \Users(C_2) |}{| \Users(C_1) \cup \Users(C_2) |}
\end{equation}
\noindent where $\Users(C)$ denotes the set of users in city $C$.
We observe that the N-CD strategy is really useful to find more users in common, however, it is not clear the actual effect this characteristic may have on the performance of CF algorithms.
At the same time, even though the P-CD strategy does not discover many users in common, since it includes much more data to train the recommenders, it may be more beneficial to some recommendation approaches.

More formally, the knowledge exploitation between cities that we propose can be done with both knowledge aggregation or linking knowledge, as we can aggregate all the cities in the source domain to make recommendations but we can also transfer knowledge (e.g., with attributes, latent features, semantic networks) between them. 
Thus, we adapt the optimization formula from \cite{DBLP:journals/tkde/AdomaviciusT05} to include both venue and cross-domain recommendation as follows:
\begin{equation}
\label{eq:adoma}
	\forall u \in \Users, l_u' = \argmax_{l \in \Items^{\Domain_T}} g(u^{\Domain_T}, l^{\Domain_T};\Users^{\Domain_S},\Items^{\Domain_S)}
\end{equation}
where $\Domain_S$ and $\Domain_T$ are the source and the target domains respectively. 
Hence, we aim to maximize the score that a user has for an item (both in the target domain) by using information about users and items in the source domain. 
	
Taking into account the characteristics shown in Section \ref{sec:cross-domain}, our cross-domain proposal satisfies:
\begin{itemize}
	\item Linked-domain recommendation, where we recommend items in the target domain (a specific city) by exploiting the knowledge derived from the source (data from other cities) and target domains. 
	\item Scenario with only user overlap, since an item (venue) from one domain will never appear in a different domain.
	\item Aggregated knowledge (merging user preferences), since we combine multiple sources of personal preferences (basically, the \checkins from various cities and the target city).
\end{itemize}

\section{Experimental setup}
\label{sec:exp}

\begin{table*}[tb]
\fontTables
\centering
\caption{Description of the temporal partition evaluated created based on the Foursquare dataset, where $U$, $I$, and $C$ denote the number of users, items, and \checkins.}
\label{t:partitions}
\vspace{-10pt}
\begin{tabular}{lrrrrrr}
\toprule
\textbf{\centering \Checkin period} & $\mathbf{\centering U}$ & $\mathbf{\centering I}$ &$ \mathbf{\centering C}$ & \textbf{\centering Density} & $\mathbf{\centering C/U}$ & $\mathbf{\centering C/I}$
\\
\midrule
\midrule
Apr'12-Sep'13 & 267k & 3.6M & 33M &  0.0034\%&  123.596 & 9.16 \\ 
\midrule
Training: May-Oct '12 & 202k & 1.1M & 4.7M & 0.0021\% & 23.267  & 4.278 \\
Test: Nov '12 & 150k & 352k & 831k & 0.0017\% & 5.540  & 2.361 \\ 
\bottomrule
\end{tabular}
\vspace{-10pt}
\end{table*}

\subsection{Dataset}
The experiments have been performed using the global-scale \checkin dataset of Foursquare\footnote{\url{https://sites.google.com/site/yangdingqi/home/foursquare-dataset}} made public by the authors of \cite{DBLP:journals/jnca/YangZCQ15,DBLP:journals/tist/YangZQ16}. 
Starting from more than $33$M \checkins, we created one temporal split containing $6$ months of data in its training split and one month for testing, more statistics are shown in Table \ref{t:partitions}.
As a pre-processing step, we performed a $2$-core before splitting the data into training and test, so that every user and item has at least $2$ preferences. 
Furthermore, in order to make a fair comparison among all the evaluated baselines, we decided to remove repetitions in a user basis, even though some algorithms, such as \irenmf, are able to exploit the frequency of users when visiting a specific venue. 

Additionally, the time of the \checkin in the original dataset was in UTC with the offset in minutes; hence, the local time is the UTC time plus the offset. 
However, we decided to work with the local date of each \checkin in order to select the training and test subsets independently of where the \checkin took place, as it is more realistic if the splits are made on the same local day for all cities, and not with respect to the global UTC schedule. 


\subsection{Methods for comparison}
\label{sec:baselines}
We test the following state-of-the-art recommenders:
\begin{itemize}
	\item Random (Rnd): random recommender.
	\item Popularity (Pop): recommender that suggests the most popular items, i.e., items with more \checkins.
	\item \AvgDist: baseline recommender that recommenders the closest POIs to the user's average location. The average is computed by calculating the midpoint of the coordinates of the POIs visited by each user.
	\item \PopGeoNN: a hybrid approach similar to the USG model proposed in~\cite{DBLP:conf/sigir/YeYLL11} that combines Pop, UB, and \AvgDist recommenders. It basically aggregates the scores of every item provided by each of the recommenders, after normalizing each score by the maximum score of each method.
	\item UB: a \knn recommender with a user-based approach~\cite{DBLP:reference/sp/NingDK15}.
	\item IB: a \knn recommender with an item-based approach~\cite{DBLP:reference/sp/NingDK15}.
	\item HKV: a matrix factorization (MF) approach as in \cite{DBLP:conf/icdm/HuKV08} that uses Alternate Least Squares in the minimization formula.
	\item \irenmf: weighted MF method proposed by \cite{DBLP:conf/cikm/LiuWSM14} and briefly described in Section \ref{sec:poi-rec}. 
	We selected this approach because, according to the comparison presented in \cite{DBLP:journals/pvldb/LiuPCY17}, \irenmf was very competitive with a lower execution time with respect to other models, such as \geomf \cite{DBLP:conf/kdd/LianZXSCR14}, \rankgeofm \cite{DBLP:conf/sigir/LiCLPK15}, or \lfbca \cite{DBLP:conf/gis/WangTM13}, which agrees with some preliminary experiments we performed in our dataset.
\end{itemize}

The parameters tested for these algorithms are detailed in Table \ref{t:parameters}, 
where $83$ different configurations were tested in total for each city.
The final parameters are shown in Table \ref{tab:parametersFoldNov}, and they were found by optimizing P@5 for the single-domain scenario. 
For every recommender except the \irenmf algorithm, we used the RankSys library~\cite{DBLP:reference/sp/CastellsHV15}; for \irenmf we used the implementation provided by \cite{DBLP:conf/cikm/LiuWSM14}, available here\footnote{\url{http://spatialkeyword.sce.ntu.edu.sg/eval-vldb17/}}. 
Source code to replicate these experiments can be found in the following Bitbucket repository: \href{https://bitbucket.org/PabloSanchezP/TempCDSeqEval}{PabloSanchezP/TempCDSeqEval}.		

\begin{table}[tb]
\normalsize
\centering
\caption{Parameters used with the evaluated recommenders. SC and SJ stand for SetCosine (as proposed in \cite{DBLP:conf/recsys/Aiolli13}, with $\alpha=0.5$) and SetJaccard, i.e., implicit versions of the well-known similarity metrics.}
\label{t:parameters}
\vspace{-10pt}
\begin{tabular}{ll}
\toprule
\textbf{Recommender} & \textbf{Parameters}                                     
\\
\midrule
\midrule
Rnd               & None                                                    \\ 
Pop           & None                                                    \\ 
{\AvgDist}           & None                                                    \\ 
\midrule
{\PopGeoNN}          & $k=100$, Similarity = SJ                                                    \\
\midrule
UB                   & \parbox{5cm}{Similarity = \{SC, SJ\}, \\$k=\{5,10, \cdots, 100\}$}        \\ 
\midrule
IB                   & \parbox{5cm}{Similarity = \{SC, SJ\}, \\$k=\{5,10, \cdots, 100\}$}        \\ 
\midrule
HKV                  & \parbox{5cm}{Factors $=\{10, 50, 100\}$, \\$\alpha=\{0.1, 1, 10\}$, $\lambda=\{0.1, 1, 10\}$} \\ 
\midrule
\irenmf               & \parbox{5cm}{$k=10$, Clusters $=50$, $\lambda_1=\lambda_2=0.015$, \\ Factors $=\{50, 100\}$, $\alpha=\{0.4, 0.6\}$, \\$\lambda_3=\{0.1,1\}$} \\ 
\bottomrule
\end{tabular}
\vspace{-10pt}
\end{table}		

\begin{table}[tb]
\fontTables
\centering
\caption{Optimal parameters of recommenders for each city. 
The order of the presented parameters is: for UB and IB, similarity and \neighhoodOR size; for HKV, factors, $\alpha$, $\lambda$; for \irenmf, factors, $\alpha$, $\lambda_3$.}
\label{tab:parametersFoldNov}
\vspace{-10pt}
\begin{tabular}{lrrrr}
\toprule
\diagbox{City}{Rec}			& UB & IB & HKV & \irenmf\\ 
\midrule
\Istanbul (\IstanbulAbbr) 
& SJ, $90$ & SC, $100$ &  $10, 10, 10$ & $100, 0.4, 1$ \\
%
\Jakarta (\JakartaAbbr) 
& SJ, $100$ & SC, $80$ & $10, 10, 10$ & $50, 0.4, 1$ \\ 
%
\KualaLumpur (\KualaLumpurAbbr) 
& SJ, $100$ & SJ, $100$ & $10, 10, 10$ & $100, 0.4, 0.1$ \\ 
%
\Mexico (\MexicoAbbr) 
& SJ, $100$ & SJ, $100$ & $10, 10, 10$ & $50, 0.4, 1$ \\ 
\midrule
\Moscow (\MoscowAbbr) 
& SC, $100$ & SJ, $100$ & $50, 10, 1$ & $100, 0.4, 1$ \\ 
%
\Santiago (\SantiagoAbbr) 
& SJ, $90$ & SJ, $80$ & $10, 10, 10$ & $100, 0.4, 1$ \\ 
%
\SaoPaulo (\SaoPauloAbbr) 
& SJ, $100$ & SJ, $100$ & $50, 10, 0.1$ & $100, 0.4, 0.1$ \\ 
%
\Tokyo (\TokyoAbbr) 
& SJ, $80$ & SC, $80$ &  $10, 10, 10$ & $100, 0.4, 1$ \\ 
%
\bottomrule
\end{tabular}
\vspace{-10pt}
\end{table}

\subsection{Evaluation methodology}
Based on the temporal split presented in Table \ref{t:partitions}, we decided to focus on the eight largest cities in terms of number of \checkins: \Mexico, \SaoPaulo, \Moscow, \KualaLumpur, \Santiago, \Tokyo, \Jakarta, and \Istanbul. 

To evaluate the recommenders, we applied the TrainItems methodology \cite{DBLP:conf/recsys/SaidB14}, where only the venues that appear in the training set of each target city are considered as candidates, except the ones already rated by the user.
Then, according to what we present in Section \ref{sec:crosspoi}, we compare the recommendations generated by the different algorithms when using different training information: either only the target city (single-domain, or SD), or a combination of cities (cross-domain, or CD) where we explore two possibilities -- nearest-based CD (N-CD) and most popular CD (P-CD).
%
Unless stated otherwise, the reported values are computed at a cutoff of $5$.

\section{Results}
In this section, we present first the results obtained when considering each city as a separate domain (Section~\ref{ss:single}) and then we analyze how different cross-domain strategies affect the performance of single-domain scenarios (Section~\ref{ss:cross}).

\begin{table*}
\caption{Performance results for the single-domain scenario in terms of nDCG@5. 
In bold, we show the highest value for each city.}
\label{t:sd}
\vspace{-10pt}
\begingroup \fontTables %
\begin {tabular}{c}%
\toprule City \\\midrule %
\IstanbulAbbr \\%
\JakartaAbbr \\%
\KualaLumpurAbbr \\%
\MexicoAbbr \\%
\midrule \MoscowAbbr \\%
\SantiagoAbbr \\%
\SaoPauloAbbr \\%
\TokyoAbbr \\\bottomrule %
\end {tabular}%
\endgroup %
\enskip
\begingroup \fontTables %
\begin {tabular}{rrrrrrrr}%
\toprule Rnd & Pop & {\AvgDist } & {\PopGeoNN } & UB & IB & HKV & {\irenmf }\\\midrule %
\pgfutilensuremath {0.000}&\pgfutilensuremath {0.054}&\pgfutilensuremath {0.001}&\pgfutilensuremath {0.067}&$\bf \pgfutilensuremath {0.073}$&\pgfutilensuremath {0.059}&\pgfutilensuremath {0.070}&\pgfutilensuremath {0.069}\\%
\pgfutilensuremath {0.000}&\pgfutilensuremath {0.066}&\pgfutilensuremath {0.001}&\pgfutilensuremath {0.067}&$\bf \pgfutilensuremath {0.070}$&\pgfutilensuremath {0.035}&\pgfutilensuremath {0.066}&\pgfutilensuremath {0.065}\\%
\pgfutilensuremath {0.000}&\pgfutilensuremath {0.066}&\pgfutilensuremath {0.001}&\pgfutilensuremath {0.070}&$\bf \pgfutilensuremath {0.073}$&\pgfutilensuremath {0.042}&\pgfutilensuremath {0.066}&\pgfutilensuremath {0.070}\\%
\pgfutilensuremath {0.000}&\pgfutilensuremath {0.041}&\pgfutilensuremath {0.001}&\pgfutilensuremath {0.043}&\pgfutilensuremath {0.044}&\pgfutilensuremath {0.013}&$\bf \pgfutilensuremath {0.047}$&\pgfutilensuremath {0.043}\\%
\midrule \pgfutilensuremath {0.000}&\pgfutilensuremath {0.027}&\pgfutilensuremath {0.002}&\pgfutilensuremath {0.032}&\pgfutilensuremath {0.037}&\pgfutilensuremath {0.017}&$\bf \pgfutilensuremath {0.039}$&\pgfutilensuremath {0.035}\\%
\pgfutilensuremath {0.000}&\pgfutilensuremath {0.051}&\pgfutilensuremath {0.001}&$\bf \pgfutilensuremath {0.054}$&\pgfutilensuremath {0.053}&\pgfutilensuremath {0.026}&\pgfutilensuremath {0.050}&\pgfutilensuremath {0.052}\\%
\pgfutilensuremath {0.000}&\pgfutilensuremath {0.053}&\pgfutilensuremath {0.001}&$\bf \pgfutilensuremath {0.057}$&\pgfutilensuremath {0.049}&\pgfutilensuremath {0.015}&\pgfutilensuremath {0.048}&\pgfutilensuremath {0.043}\\%
\pgfutilensuremath {0.000}&\pgfutilensuremath {0.069}&\pgfutilensuremath {0.001}&$\bf \pgfutilensuremath {0.070}$&\pgfutilensuremath {0.069}&\pgfutilensuremath {0.048}&\pgfutilensuremath {0.059}&\pgfutilensuremath {0.068}\\\bottomrule %
\end {tabular}%
\endgroup %
\end{table*}

\begin{table*}
\caption{Performance results for the cross-domain scenario in terms of nDCG@5: N-CD and P-CD indicate a cross-domain strategy using the nearest $7$ cities and the most popular $7$ cities, respectively. 
The improvement in performance with respect to the single-domain scenario (Table~\ref{t:sd}) is represented as $\Delta$(\%), where $\blacktriangleSmall$ ($\blacktriangledownSmall$) denotes the largest positive (negative) improvement.
}
\label{t:cd}
\vspace{-10pt}
\begingroup \fontTables %
\begin {tabular}{rr}%
\toprule City \\\midrule %
\multirow {4}{*}{\IstanbulAbbr }&N-CD\\%
&$\Delta $(\%)\\%
&P-CD\\%
&$\Delta $(\%)\\%
\midrule \multirow {4}{*}{\JakartaAbbr }&N-CD\\%
&$\Delta $(\%)\\%
&P-CD\\%
&$\Delta $(\%)\\%
\midrule \multirow {4}{*}{\KualaLumpurAbbr }&N-CD\\%
&$\Delta $(\%)\\%
&P-CD\\%
&$\Delta $(\%)\\%
\midrule \multirow {4}{*}{\MexicoAbbr }&N-CD\\%
&$\Delta $(\%)\\%
&P-CD\\%
&$\Delta $(\%)\\%
\midrule \multirow {4}{*}{\MoscowAbbr }&N-CD\\%
&$\Delta $(\%)\\%
&P-CD\\%
&$\Delta $(\%)\\%
\midrule \multirow {4}{*}{\SantiagoAbbr }&N-CD\\%
&$\Delta $(\%)\\%
&P-CD\\%
&$\Delta $(\%)\\%
\midrule \multirow {4}{*}{\SaoPauloAbbr }&N-CD\\%
&$\Delta $(\%)\\%
&P-CD\\%
&$\Delta $(\%)\\%
\midrule \multirow {4}{*}{\TokyoAbbr }&N-CD\\%
&$\Delta $(\%)\\%
&P-CD\\%
&$\Delta $(\%)\\\bottomrule %
\end {tabular}%
\endgroup %
\enskip
\begingroup \fontTables %
\begin {tabular}{rrrrrr}%
\toprule {\AvgDist } & {\PopGeoNN } & UB & IB & HKV & {\irenmf }\\\midrule %
\pgfutilensuremath {0.001}&\pgfutilensuremath {0.068}&$\bf \pgfutilensuremath {0.073}$&\pgfutilensuremath {0.057}&\pgfutilensuremath {0.071}&\pgfutilensuremath {0.059}\\%
\textit {\pgfmathprintnumber [fixed,precision=1,fixed zerofill]{-9.72512404444869}}&\textit {\pgfmathprintnumber [fixed,precision=1,fixed zerofill]{1.6145212594426}}&\textit {\pgfmathprintnumber [fixed,precision=1,fixed zerofill]{0.318953468586154}}&\textit {\pgfmathprintnumber [fixed,precision=1,fixed zerofill]{-3.20742918712793}}&$\blacktriangleSmall $\textit {\pgfmathprintnumber [fixed,precision=1,fixed zerofill]{1.98417729021516}}&$\blacktriangledownSmall $\textit {\pgfmathprintnumber [fixed,precision=1,fixed zerofill]{-14.8179300472206}}\\%
\pgfutilensuremath {0.001}&\pgfutilensuremath {0.068}&$\bf \pgfutilensuremath {0.073}$&\pgfutilensuremath {0.059}&\pgfutilensuremath {0.068}&\pgfutilensuremath {0.052}\\%
\textit {\pgfmathprintnumber [fixed,precision=1,fixed zerofill]{-0.103855640659491}}&$\blacktriangleSmall $\textit {\pgfmathprintnumber [fixed,precision=1,fixed zerofill]{0.911787579179982}}&\textit {\pgfmathprintnumber [fixed,precision=1,fixed zerofill]{0.3521646220307}}&\textit {\pgfmathprintnumber [fixed,precision=1,fixed zerofill]{-0.042588174064496}}&\textit {\pgfmathprintnumber [fixed,precision=1,fixed zerofill]{-3.40912280498949}}&$\blacktriangledownSmall $\textit {\pgfmathprintnumber [fixed,precision=1,fixed zerofill]{-24.7427407214556}}\\%
\midrule \pgfutilensuremath {0.001}&\pgfutilensuremath {0.070}&$\bf \pgfutilensuremath {0.075}$&\pgfutilensuremath {0.034}&\pgfutilensuremath {0.070}&\pgfutilensuremath {0.062}\\%
$\blacktriangledownSmall $\textit {\pgfmathprintnumber [fixed,precision=1,fixed zerofill]{-17.251464228136}}&\textit {\pgfmathprintnumber [fixed,precision=1,fixed zerofill]{4.75093754686967}}&$\blacktriangleSmall $\textit {\pgfmathprintnumber [fixed,precision=1,fixed zerofill]{6.67945859631261}}&\textit {\pgfmathprintnumber [fixed,precision=1,fixed zerofill]{-3.94746905270155}}&\textit {\pgfmathprintnumber [fixed,precision=1,fixed zerofill]{6.31356274351299}}&\textit {\pgfmathprintnumber [fixed,precision=1,fixed zerofill]{-4.51502595021422}}\\%
\pgfutilensuremath {0.001}&\pgfutilensuremath {0.068}&$\bf \pgfutilensuremath {0.071}$&\pgfutilensuremath {0.035}&\pgfutilensuremath {0.060}&\pgfutilensuremath {0.058}\\%
\textit {\pgfmathprintnumber [fixed,precision=1,fixed zerofill]{-9.88711635005392}}&$\blacktriangleSmall $\textit {\pgfmathprintnumber [fixed,precision=1,fixed zerofill]{0.63295912407905}}&\textit {\pgfmathprintnumber [fixed,precision=1,fixed zerofill]{0.366280909932099}}&\textit {\pgfmathprintnumber [fixed,precision=1,fixed zerofill]{-0.803854858800172}}&\textit {\pgfmathprintnumber [fixed,precision=1,fixed zerofill]{-8.50497980610302}}&$\blacktriangledownSmall $\textit {\pgfmathprintnumber [fixed,precision=1,fixed zerofill]{-10.6515826994838}}\\%
\midrule \pgfutilensuremath {0.001}&\pgfutilensuremath {0.072}&$\bf \pgfutilensuremath {0.076}$&\pgfutilensuremath {0.040}&\pgfutilensuremath {0.075}&\pgfutilensuremath {0.069}\\%
$\blacktriangledownSmall $\textit {\pgfmathprintnumber [fixed,precision=1,fixed zerofill]{-36.2732137660558}}&\textit {\pgfmathprintnumber [fixed,precision=1,fixed zerofill]{2.12437363377592}}&\textit {\pgfmathprintnumber [fixed,precision=1,fixed zerofill]{4.13194103757107}}&\textit {\pgfmathprintnumber [fixed,precision=1,fixed zerofill]{-3.87005200392921}}&$\blacktriangleSmall $\textit {\pgfmathprintnumber [fixed,precision=1,fixed zerofill]{13.7875434494312}}&\textit {\pgfmathprintnumber [fixed,precision=1,fixed zerofill]{-0.979369738199287}}\\%
\pgfutilensuremath {0.001}&\pgfutilensuremath {0.070}&$\bf \pgfutilensuremath {0.073}$&\pgfutilensuremath {0.042}&\pgfutilensuremath {0.065}&\pgfutilensuremath {0.064}\\%
\textit {\pgfmathprintnumber [fixed,precision=1,fixed zerofill]{-0.760324866079168}}&\textit {\pgfmathprintnumber [fixed,precision=1,fixed zerofill]{0.019114184721242}}&$\blacktriangleSmall $\textit {\pgfmathprintnumber [fixed,precision=1,fixed zerofill]{0.292651714453952}}&\textit {\pgfmathprintnumber [fixed,precision=1,fixed zerofill]{-0.313082314000179}}&\textit {\pgfmathprintnumber [fixed,precision=1,fixed zerofill]{-1.55410156152136}}&$\blacktriangledownSmall $\textit {\pgfmathprintnumber [fixed,precision=1,fixed zerofill]{-8.75781269072213}}\\%
\midrule \pgfutilensuremath {0.001}&\pgfutilensuremath {0.044}&\pgfutilensuremath {0.045}&\pgfutilensuremath {0.013}&$\bf \pgfutilensuremath {0.045}$&\pgfutilensuremath {0.040}\\%
$\blacktriangleSmall $\textit {\pgfmathprintnumber [fixed,precision=1,fixed zerofill]{13.2741838153101}}&\textit {\pgfmathprintnumber [fixed,precision=1,fixed zerofill]{2.20817630654149}}&\textit {\pgfmathprintnumber [fixed,precision=1,fixed zerofill]{1.61627051771237}}&\textit {\pgfmathprintnumber [fixed,precision=1,fixed zerofill]{-6.46216295124836}}&\textit {\pgfmathprintnumber [fixed,precision=1,fixed zerofill]{-5.03259772350213}}&$\blacktriangledownSmall $\textit {\pgfmathprintnumber [fixed,precision=1,fixed zerofill]{-6.79664482176526}}\\%
\pgfutilensuremath {0.001}&\pgfutilensuremath {0.044}&$\bf \pgfutilensuremath {0.045}$&\pgfutilensuremath {0.013}&\pgfutilensuremath {0.037}&\pgfutilensuremath {0.037}\\%
\textit {\pgfmathprintnumber [fixed,precision=1,fixed zerofill]{-0.21308331557636}}&$\blacktriangleSmall $\textit {\pgfmathprintnumber [fixed,precision=1,fixed zerofill]{1.32814289991275}}&\textit {\pgfmathprintnumber [fixed,precision=1,fixed zerofill]{1.23572289117144}}&\textit {\pgfmathprintnumber [fixed,precision=1,fixed zerofill]{-0.064171122994926}}&$\blacktriangledownSmall $\textit {\pgfmathprintnumber [fixed,precision=1,fixed zerofill]{-22.0711379849638}}&\textit {\pgfmathprintnumber [fixed,precision=1,fixed zerofill]{-13.6123570974654}}\\%
\midrule \pgfutilensuremath {0.002}&\pgfutilensuremath {0.033}&\pgfutilensuremath {0.038}&\pgfutilensuremath {0.017}&$\bf \pgfutilensuremath {0.040}$&\pgfutilensuremath {0.034}\\%
$\blacktriangledownSmall $\textit {\pgfmathprintnumber [fixed,precision=1,fixed zerofill]{-6.8894290380804}}&\textit {\pgfmathprintnumber [fixed,precision=1,fixed zerofill]{0.778732642225716}}&\textit {\pgfmathprintnumber [fixed,precision=1,fixed zerofill]{2.52736749350336}}&\textit {\pgfmathprintnumber [fixed,precision=1,fixed zerofill]{-0.740620834002244}}&$\blacktriangleSmall $\textit {\pgfmathprintnumber [fixed,precision=1,fixed zerofill]{3.32744795225745}}&\textit {\pgfmathprintnumber [fixed,precision=1,fixed zerofill]{-1.14446569291178}}\\%
\pgfutilensuremath {0.002}&\pgfutilensuremath {0.032}&$\bf \pgfutilensuremath {0.037}$&\pgfutilensuremath {0.018}&\pgfutilensuremath {0.036}&\pgfutilensuremath {0.029}\\%
\textit {\pgfmathprintnumber [fixed,precision=1,fixed zerofill]{-0.59125964010255}}&\textit {\pgfmathprintnumber [fixed,precision=1,fixed zerofill]{0.058678053775465}}&\textit {\pgfmathprintnumber [fixed,precision=1,fixed zerofill]{0.334577696195088}}&$\blacktriangleSmall $\textit {\pgfmathprintnumber [fixed,precision=1,fixed zerofill]{1.12386233220724}}&\textit {\pgfmathprintnumber [fixed,precision=1,fixed zerofill]{-7.69827780846584}}&$\blacktriangledownSmall $\textit {\pgfmathprintnumber [fixed,precision=1,fixed zerofill]{-17.3538238931287}}\\%
\midrule \pgfutilensuremath {0.001}&\pgfutilensuremath {0.059}&$\bf \pgfutilensuremath {0.060}$&\pgfutilensuremath {0.025}&\pgfutilensuremath {0.060}&\pgfutilensuremath {0.055}\\%
\textit {\pgfmathprintnumber [fixed,precision=1,fixed zerofill]{8.04488759849318}}&\textit {\pgfmathprintnumber [fixed,precision=1,fixed zerofill]{8.68172013268209}}&\textit {\pgfmathprintnumber [fixed,precision=1,fixed zerofill]{12.9803027164287}}&$\blacktriangledownSmall $\textit {\pgfmathprintnumber [fixed,precision=1,fixed zerofill]{-5.00971630669978}}&$\blacktriangleSmall $\textit {\pgfmathprintnumber [fixed,precision=1,fixed zerofill]{20.2070444424755}}&\textit {\pgfmathprintnumber [fixed,precision=1,fixed zerofill]{4.9195079569325}}\\%
\pgfutilensuremath {0.001}&$\bf \pgfutilensuremath {0.055}$&\pgfutilensuremath {0.054}&\pgfutilensuremath {0.026}&\pgfutilensuremath {0.046}&\pgfutilensuremath {0.045}\\%
\textit {\pgfmathprintnumber [fixed,precision=1,fixed zerofill]{-0.711837595570819}}&$\blacktriangleSmall $\textit {\pgfmathprintnumber [fixed,precision=1,fixed zerofill]{1.13705794015841}}&\textit {\pgfmathprintnumber [fixed,precision=1,fixed zerofill]{0.86342401517716}}&\textit {\pgfmathprintnumber [fixed,precision=1,fixed zerofill]{-0.265463233342117}}&\textit {\pgfmathprintnumber [fixed,precision=1,fixed zerofill]{-7.86662592793351}}&$\blacktriangledownSmall $\textit {\pgfmathprintnumber [fixed,precision=1,fixed zerofill]{-13.2295696830722}}\\%
\midrule \pgfutilensuremath {0.001}&$\bf \pgfutilensuremath {0.057}$&\pgfutilensuremath {0.056}&\pgfutilensuremath {0.016}&\pgfutilensuremath {0.056}&\pgfutilensuremath {0.046}\\%
$\blacktriangledownSmall $\textit {\pgfmathprintnumber [fixed,precision=1,fixed zerofill]{-7.11109447166925}}&\textit {\pgfmathprintnumber [fixed,precision=1,fixed zerofill]{0.410523019512442}}&$\blacktriangleSmall $\textit {\pgfmathprintnumber [fixed,precision=1,fixed zerofill]{15.4210637065641}}&\textit {\pgfmathprintnumber [fixed,precision=1,fixed zerofill]{5.51455246422059}}&\textit {\pgfmathprintnumber [fixed,precision=1,fixed zerofill]{15.2331715530803}}&\textit {\pgfmathprintnumber [fixed,precision=1,fixed zerofill]{7.32489526061695}}\\%
\pgfutilensuremath {0.001}&$\bf \pgfutilensuremath {0.057}$&\pgfutilensuremath {0.049}&\pgfutilensuremath {0.015}&\pgfutilensuremath {0.047}&\pgfutilensuremath {0.034}\\%
\textit {\pgfmathprintnumber [fixed,precision=1,fixed zerofill]{-9.24096179340672}}&$\blacktriangleSmall $\textit {\pgfmathprintnumber [fixed,precision=1,fixed zerofill]{0.533234912727497}}&\textit {\pgfmathprintnumber [fixed,precision=1,fixed zerofill]{-0.222992356040803}}&\textit {\pgfmathprintnumber [fixed,precision=1,fixed zerofill]{-0.240168117682422}}&\textit {\pgfmathprintnumber [fixed,precision=1,fixed zerofill]{-2.08981279951547}}&$\blacktriangledownSmall $\textit {\pgfmathprintnumber [fixed,precision=1,fixed zerofill]{-20.1877995507803}}\\%
\midrule \pgfutilensuremath {0.000}&$\bf \pgfutilensuremath {0.073}$&\pgfutilensuremath {0.073}&\pgfutilensuremath {0.048}&\pgfutilensuremath {0.064}&\pgfutilensuremath {0.071}\\%
$\blacktriangledownSmall $\textit {\pgfmathprintnumber [fixed,precision=1,fixed zerofill]{-15.5840793316293}}&\textit {\pgfmathprintnumber [fixed,precision=1,fixed zerofill]{4.89370078477253}}&\textit {\pgfmathprintnumber [fixed,precision=1,fixed zerofill]{5.37507103737261}}&\textit {\pgfmathprintnumber [fixed,precision=1,fixed zerofill]{-0.220671355238621}}&$\blacktriangleSmall $\textit {\pgfmathprintnumber [fixed,precision=1,fixed zerofill]{8.66964388184181}}&\textit {\pgfmathprintnumber [fixed,precision=1,fixed zerofill]{4.17523950527802}}\\%
\pgfutilensuremath {0.001}&$\bf \pgfutilensuremath {0.070}$&\pgfutilensuremath {0.069}&\pgfutilensuremath {0.048}&\pgfutilensuremath {0.064}&\pgfutilensuremath {0.064}\\%
\textit {\pgfmathprintnumber [fixed,precision=1,fixed zerofill]{-0.28659160696005}}&\textit {\pgfmathprintnumber [fixed,precision=1,fixed zerofill]{-0.164841748494085}}&\textit {\pgfmathprintnumber [fixed,precision=1,fixed zerofill]{-0.199692314451161}}&\textit {\pgfmathprintnumber [fixed,precision=1,fixed zerofill]{-0.102774922918878}}&$\blacktriangleSmall $\textit {\pgfmathprintnumber [fixed,precision=1,fixed zerofill]{8.56598651538132}}&$\blacktriangledownSmall $\textit {\pgfmathprintnumber [fixed,precision=1,fixed zerofill]{-6.087792532385}}\\\bottomrule %
\end {tabular}%
\endgroup %
\end{table*}

\subsection{Analysis of single-domain performance}
\label{ss:single}
We show in Table \ref{t:sd} the results for the recommenders presented in Section~\ref{sec:baselines} in terms of nDCG@5 for the single-domain scenario. 
The first thing we note is the low values obtained by the recommenders. This is mostly due to the high sparsity of the data (see Table~\ref{t:partitions}), together with the fact that we are using a temporal split, which makes the recommendation task even more difficult, since, for instance, some of the few items a user may have in her test set may not appear in the training set at all, making such recommendation impossible to be produced by a collaborative algorithm. 

Moreover, the \AvgDist recommender is the second worst algorithm (after Rnd), which could be attributed to how we processed the data: since each venue visited by the user has now the same importance (because we removed repetitions), the user's center cannot consider the most frequent venues in their computation and, hence, it might not represent her preferred areas in the city.

Similarly, we observe that the \irenmf approach, even though it remains very competitive, it never emerges as the optimal recommender. 
This might be caused by several reasons; first, as with the \AvgDist recommender, it is able to work with repeated values and without them it loses some efficiency; secondly, its claimed superior performance was only tested using a random split in~\cite{DBLP:conf/cikm/LiuWSM14} instead of a temporal evaluation; and, finally, classical recommendation algorithms such as Pop or standard CF approaches were neglected in \cite{DBLP:journals/pvldb/LiuPCY17} which, together with our previous discussion, definitely disturbs such comparisons.

For the rest of the recommenders, we observe that UB is one of the best approaches for most of the cities. 
This result might be misleading as this recommender actually has less user coverage than other approaches~\cite{DBLP:reference/sp/GunawardanaS15}. 
It is interesting to note the relatively high performance of \PopGeoNN, since it is able to beat the rest of the baselines in many cities, despite its simplicity and the fact that we did not perform any parameter tuning.

\subsection{Performance comparison of cross-domain strategies}
\label{ss:cross}

Table \ref{t:cd} shows the results for all cities with two different cross-domain approaches: P-CD (for every city, the training set is built by combining the training data from the eight selected cities) and N-CD (the training set is built by taking the nearest $7$ cities with respect to the target city, so the number of cities under consideration is comparable to that of P-CD).
In the same table, we show the improvement with respect to the single-domain scenario for every algorithm except Rnd and Pop, since they do not change their recommendations when changing the training set and keeping the same test set.
 
We observe that the performance improvement when using the P-CD strategy is usually negligible, except for \TokyoAbbr and HKV; in general, most of the improvements when using this strategy are very close to zero and, for many of the city-recommender combinations, extremely negative.
This is especially true for the MF approaches (HKV and \irenmf), which seems to indicate that having more data available so that the sparsity is reduced does not guarantee better recommendations.
Moreover, N-CD usually produces larger improvements with less training data involved (since the nearest cities always include less \checkins than the originally selected cities, which were the most popular ones in our dataset).
This seems to confirm that better data is more useful than more data, in particular, the amount of user overlap is a good signal of the impact that a cross-domain strategy may obtain (as evidenced by the data presented before in Table~\ref{t:usersAverage}).

Classical CF algorithms like UB and HKV are able to exploit more successfully the information coming from the source domains, except, as discussed before, HKV with P-CD; the IB approach, on the other hand, obtained in general very poor results.
We argue this trend for the CF techniques is related to whether the new users -- recall that the new items will never have overlap with the target items, since this type of cross-domain has only user overlap by definition -- have some kind of interaction with the target city.
In this sense, when combining information from nearby cities it is more likely to find similar users with useful suggestions or learning relevant latent representations more related to the target items.

On the other hand, cross-domain strategies tend to deteriorate the performance of the techniques based on geographical distances (\AvgDist and \irenmf). The reason for this might be quite obvious, since considering other cities to compute a new user's centroid will certainly move such centroid far away from the target city, which is not useful when we are only interested in recommending venues inside of that specific city.

Therefore, we are able to answer our two research questions. 
First, regarding \textbf{RQ1} (Are state-of-the-art recommenders able to exploit cross-domain information in the context of venue recommendation?) we have seen that some classic recommenders are able to benefit from cross-domain information if the cities are selected properly, however, other approaches more tailored for venue recommendation (such as those exploiting geographical information) seem to decrease their performance when exploiting knowledge from other domains.
Second, as an answer to \textbf{RQ2} (Which is the best cross-domain strategy in terms of relevance?), we can conclude that selecting cities by proximity (the closest ones) has a greater benefit than selecting them by the amount of information they contain (popularity).

\subsection{Discussion}

According to the results obtained, applying cross-domain techniques can improve the results obtained in some situations, although their effect is not as great as one might expect (mainly due to the temporal split we applied and the sparsity of the dataset). 
However, since we mostly explored very basic recommendation techniques, these results are very promising. 
First of all, because we have seen that some of the algorithms are able to make better recommendations (in some cases, up to a $20\%$ improvement) -- however, further analysis should be done to understand the impact of such improvement in other evaluation dimensions, such as novelty or diversity. 
Secondly, due to the well-known popularity bias~\cite{DBLP:journals/umuai/JannachLKJ15}, such a simple technique could outperform other methods like IB or \AvgDist (see Table~\ref{t:sd}), even though this type of baseline is usually ignored in POI recommendation literature.
Finally, a negative result we observed is that when the distance between the venues is considered, applying cross-domain can be counterproductive and is not recommended.

It should be noted that our results are consistent with those discussed in \cite{DBLP:conf/recsys/SahebiB15}, where the authors found that there are specific experiments in which cross-domain recommendation works worse than classic single-domain recommendations, even though in general it behaves better or as good as single-domain strategies. 
Nonetheless, only comparisons between single-domain and cross-domain approaches on three different algorithms and without considering any temporal split were presented in that paper; hence, our work helps on generalizing the conclusions obtained in such paper.

We want to emphasize that considering knowledge from different cities (understood as different domains), despite being computationally more expensive, has a clear advantage: such system would only need to train once for each city where recommendations are required, whereas considering each city as an isolated training domain (called single-domain in this paper) only allows to generate recommendations for a single city; hence, the recommendation model built can be re-used more often in the former case, at the expense of being more expensive in terms of memory and time consuming.
However, as we have shown here, if the cross-domain model is generated based on the \textit{right} cities, significant performance improvements can be achieved, not always by selecting the cities with more information but those that are closer and more likely to have overlap in their users, probably because they are culturally related and share similar mobility patterns~\cite{DBLP:journals/tist/YangZQ16}.


\section{Related work}
\label{sec:soa}
Cross-domain techniques can help recommendation systems in a number of situations. 
Specifically, travel recommendation was addressed as a potential target of these techniques \citep{DBLP:journals/tbd/Zheng15}. There have been some papers in which researchers explored different ways to combine sources of information to be applied in tourism. 
For example, in \citep{DBLP:journals/jitt/SabouOBS16} the authors describe TourMIS, a dataset of European statistics of tourism data, where they show the usefulness of combining different data sources (economy, tourism, and sustainability) to make relations between them. 
Although they indicate that sometimes it is difficult to integrate sources from very different domains, most of those problems can be solved using Linked Data approaches. 
Nevertheless, the dataset described was not used to produce venue recommendations, only for statistical analysis. 
Another approach of cross-domain in tourism can be seen in \citep{TobiasSemanticCross}, where the authors did not recommend POIs but music artists depending on the monument the user is visiting; they do this by building a semantic network of venues and artists and the relations between them. 
In \cite{DBLP:conf/www/ZhengZXY10}, the authors exploited information from different sources like location-activity, location-feature, and activity-activity correlations to enhance the performance, showing a $20\%$ improvement over a basic algorithm that does not use any additional data.  

However, the most similar approach to this paper that we found is \cite{DBLP:journals/kais/ZhangW16}. 
In that article, the authors perform a so-called cross-region recommendation, and considered each region as a different domain.
One important difference with our work is that, whereas a division by cities is natural, they used regions computed by performing clustering on the venues.
Additionally, the datasets used (Foursquare and Yelp) are smaller than the one reported here and no temporal evaluation was performed, only a standard cross validation methodology. 
Hence, our paper offers a complementary view on a related problem, from a more realistic perspective (since we explicitly address a time-aware evaluation) with a larger dataset and taking into account the \checkins in different cities to perform the recommendations.

\section{Conclusions and future work}
\label{sec:conclusions}
In this paper we have explored venue recommendation with a novel approach, by applying concepts of cross-domain recommendation under a realistic scenario, by using a temporal evaluation methodology. 
We have shown an empirical evaluation comparing the performance of state-of-the-art recommenders under different settings (single-domain and cross-domain applied on the closest cities and the most popular ones). 
Even though the behavior varies in every city, the cross-domain strategy based on the closest cities tends to produce better results; in the future, we would like to exploit this observation to create a generic recommender system for venue suggestion that takes this information into account, since so far in this work we have not proposed any new algorithm using cross-domain data.

In any case, we believe there is still room for improvement. 
First of all, the evaluated algorithms are very simple, this study should probably be repeated considering more complex methods such as \cite{DBLP:conf/cikm/ManotumruksaMO17} or \cite{DBLP:conf/recsys/GaoTHL13}. Nevertheless, POI recommendation algorithms that exploit geographic information may be negatively affected by the cross-domain strategies as user movement patterns may be modified if we use information from other cities checkins. On the other hand, the temporal evaluation methodology should also be analyzed more carefully, especially regarding the effect of seasonal trends and how it may affect the knowledge transfer techniques (since not enough interactions or users might be available or active at the same time).

We aim to further extend how the cities are selected by computing some kind of similarity between them using content and cultural information~\cite{DBLP:journals/tist/YangZQ16}.
We would also like to exploit the venue categories, social connections, or other content information, and see how that information is affected when using cross-domain strategies.
Furthermore, due to the high sparsity of this recommendation task, we believe that a study to analyze the cold-start problem is needed in order to check if cross-domain strategies perform well under these circumstances. Even though in this paper we performed a $2$-core pre-processing and this setting is still realistic enough for several situations, a more thorough analysis should be performed to understand the sensitivity of this problem to the different algorithms and cross-domain strategies.

\begin{acks}
This work was funded by the research project \grantnum{1}{TIN2016-80630-P} (\grantsponsor{1}{MINECO}{}).
\end{acks}


\bibliographystyle{ACM-Reference-Format}


\end{document}